\documentclass[prl, twocolumn]{revtex4}
\usepackage{amssymb}
\usepackage{graphicx}
\usepackage{amsmath}
\usepackage{amsbsy}
\usepackage{amsthm}
\usepackage{bbm}
\usepackage{bm}
\newcommand{\id}{\mathbbm{1}}
\newcommand{\R}{\mathbbm{R}}

\newcommand{\N}{\mathbbm{N}}
\newcommand{\tr}{\textnormal{Tr}}
\newcommand{\la}{\langle}
\newcommand{\ra}{\rangle}

\begin{document}

\title{Multipartite entanglement detection via structure factors}

\author{Philipp Krammer}
\affiliation{Faculty of Physics, University of Vienna, A-1090 Vienna, Austria}
\author{Hermann Kampermann}
\affiliation{Institut f{\"u}r Theoretische Physik III, Heinrich-Heine-Universit{\"a}t D{\"u}sseldorf, D-40225 D{\"u}sseldorf, Germany}
\author{Dagmar Bru{\ss}}
\affiliation{Institut f{\"u}r Theoretische Physik III, Heinrich-Heine-Universit{\"a}t D{\"u}sseldorf, D-40225 D{\"u}sseldorf, Germany}
\author{Reinhold A. Bertlmann}
\affiliation{Faculty of Physics, University of Vienna, A-1090 Vienna, Austria}
\author{Leong Chuang Kwek}
\affiliation{National Institute of Education and Institute of Advanced Studies,
Nanyang Technological University}
\affiliation{Centre for Quantum Technologies, National University of Singapore, Singapore 117543}
\author{Chiara Macchiavello}
\affiliation{Dipartimento di Fisica ``A. Volta'' and INFN-Sezione di Pavia, University of Pavia, I-27100 Pavia, Italy}

\begin{abstract}
We establish a relation between entanglement of a many-body system and its diffractive properties, where the link
is given by structure factors.
Based on these, we provide a general analytical construction of multi-qubit entanglement witnesses.
The proposed witnesses contain two-point correlations. They could be either measured in a scattering experiment
or via local measurements, depending on the underlying physical system.
For some explicit examples of witnesses we analyze the properties of the states that are detected by them.
We further study the robustness of these witnesses with respect to noise.
\end{abstract}

\maketitle

Multipartite entanglement is a remarkable property of quantum systems that outlines fundamental discrepancies
to classical physics and occurs at the interface between quantum information and many-body physics.
A useful tool to verify the presence of multipartite entanglement are entanglement witnesses.
Their experimental implementations via local measurements
are presented, e.g., in Refs.~\cite{bourennane04, barbieri03, altepeter05, kiesel07, prevedel09, wieczorek09}.
Here we introduce a method to detect multipartite entanglement also in a scattering experiment, when
in general no local access to the individual subsystems is possible.

In this paper we present a construction for multipartite entanglement witnesses based on linear
combinations of operators associated with structure factors.
Structure factors consist of two-point correlations, and have been widely used in
condensed-matter
physics. They are measurable in scattering experiments, e.g. via neutron
scattering in condensed-matter systems, or via light scattering off optical lattices.
For example, structure factors are employed in experiments
\cite{tennant03, huberman05} that investigate the crystal structures of particular molecules that
can be described via Hamiltionians of spin chains. One-particle spectral functions are investigated for various
materials in \cite{kampf90, suck07, kondo09}.
Also, much theoretical work on determining the
structure factor and the related spectral weights of particular spin chain models has been done, numerically and
analytically \cite{barnes99, hamer03, zheng03, hamer06, pereira06}. Moreover, the structure factor plays an important
role in the physics of atoms in optical lattices since it is related to the visibility of the
interference pattern \cite{gerbier05}. Impurity contributions to the static structure factors have been
predicted in \cite{eggert95}.
In this paper we show how witness expectation values can be determined from global diffractive properties,
via the corresponding structure factors. Therefore, we establish a link between
multipartite entanglement and diffractive properties of many-body systems, based on two-point correlations.
Let us note that expectation values of our witnesses can be also
measured locally \cite{guehne02}
in any suitable physical system where individual subsystems are accessible, e.g. entangled
photons.

We consider a multipartite quantum system consisting of $N$ subsystems.
The construction of our entanglement witnesses is based on the operator
$\hat{S}^{\alpha \beta}(k) := \sum_{i<j} e^{ik(r_i -r_j)} S_i^{\alpha} S_j^\beta$,
where $i,j$ denote the $i$-th and $j$-th spins, $r_i, r_j$ their positions, and
$\alpha,\beta = x,y,z$. Here, $S_i^\alpha$ are the spin operators and $k$ is the wave-vector
transfer, where we consider the one dimensional situation. The expectation value of this operator is  the static
structure factor (apart from a different summation and normalization, this form appears in, e.g.,
Ref.~\cite{hamer06})
\begin{equation} \label{statstrucfac}
    S^{\alpha \beta} (k) = \sum_{i<j} e^{ik(r_j -r_i)} \la S_i^{\alpha} S_j^\beta \ra \,,
\end{equation}
which is a function of $k$. The dynamic formulation of the structure factor describes the
``response'' of the probed system \cite{marshall71}. Below, we will normalize the distance between two
neighboring spins to one. The distance between spins is either defined via the periodic structure, or - in a
non-periodic situation, like e.g. for entangled photons - via the labels of the spins.
So far we have considered general spin operators. In the following we will focus on the case of spin 1/2,
where $S^\alpha$ corresponds to the Pauli operator $\sigma^\alpha$.

An entanglement witness is defined as a Hermitian operator $W$ that detects the entanglement of a state $\rho$
if it has a negative expectation value for this state, $\la W \ra_\rho = \tr (\rho W) < 0$ while at the same
time
$\tr (\sigma W) \geq 0$ for all separable states $\sigma$ \cite{horodecki96, terhal00}.
We construct a class of entanglement witnesses using the static structure factor operator as
\begin{equation} \label{genentwit}
	W(k) := \id_N - \Sigma(k) \,,
\end{equation}
where $\id_N$ is the identity operator on the $2^N$-dimensional Hilbert space and
\begin{equation} \label{sigmagen1}
	\Sigma(k) = \frac{1}{2} [\bar\Sigma(k)+\bar\Sigma(-k)]
\end{equation}
with
\begin{align} \label{sigmagen}
	\bar\Sigma(k) &= \frac{1}{B(N,2)} \left( c_x \hat{S}^{xx}(k) + c_y \hat{S}^{yy}(k) + c_z \hat{S}^{zz}(k) \right), \nonumber\displaybreak[0]\\
	&\quad c_i \in \R, \ |c_i| \leq 1.
\end{align}
Here $B(N,2)$ is the binomial coefficient. The restriction to real coefficients $c_i$
ensures that the witness operator
is Hermitian, while the choice $|c_i|\leq 1$ ensures normalization.
The meaning of the parameter $k$ depends on the physical system: for the detection of entanglement for states of multiple photons, it just fixes a sign rule for the two-point correlation terms of the witness and has no further physical meaning. For spin chains, as mentioned, $k$ is the wave vector transfer in scattering experiments. Here the entanglement witness can be determined via a readout of values of the structure factor for various $k$. The expectation value of the witness is $\la W(k) \ra = 1 - \la \Sigma(k) \ra$. The crucial point is to show that for all product states $\rho_p^N := \rho_1 \otimes \rho_2 \otimes \ldots \otimes \rho_N$ we have $\la W(k) \ra_{\rho_p^N} \geq 0$ (due to convexity the same is true for separable states that are a convex combination of projectors onto product states). It is helpful to remember the Bloch vector form of a qubit state,
\begin{equation} \label{blochqubit}
	\rho_i = \frac{1}{2} \left( \id + \sum_\alpha n^i_\alpha \sigma^\alpha \right) \,, (n_x^i)^2 +(n^i_y)^2 +(n_z^i)^2 \leq 1 \,.
\end{equation}

The product state $\rho_p^N$ should then be regarded as a product of Bloch decompositions
$\eqref{blochqubit}$. In this way we obtain the following bound on the expectation value of
$\Sigma(k)$ for product states:
\begin{align} \label{proofprod}
	| \la &\Sigma(k) \ra_{\rho_p^N} | = \nonumber\\
	&=\frac{1}{B(N,2)} | \sum_{i<j} \Big( c_x \cos{(km_{ij})} \la \sigma^x_i \sigma^x_j \ra
\nonumber\displaybreak[0]\\
	& \qquad \qquad + c_y \cos{(km_{ij})} \la \sigma^y_i \sigma^y_j \ra + c_z \cos{(km_{ij})} \la \sigma^z_i \sigma^z_j \ra \Big)
| \nonumber\displaybreak[0]\\
	&= \frac{1}{B(N,2)} | \sum_{i<j} \Big( c_x \cos{(km_{ij})} n_x^i n_x^j
\nonumber\\
	& \qquad \qquad + c_y \cos{(km_{ij})} n_y^i n_y^j + c_z \cos{(km_{ij})} n_z^i n_z^j \Big)| \nonumber\displaybreak[0]\\
	&\leq \frac{1}{B(N,2)} \sum_{i<j} \left( |n_x^i| |n_x^j| + |n_y^i| |n_y^j|+ |n_z^i| |n_z^j| \right) \leq 1 \,,
\end{align}
where $m_{ij} = r_j - r_i \in \N$. Thus, for product states we have $\la W(k) \ra_{\rho_p^N} \geq 0$.

%
Which states can be detected by $W(k)$? For $k=0$ symmetric states like the Dicke states can be detected. These states are defined as pure states that are a superposition of all possible permutations of $l$ excitations (states $|1 \ra$) in $N$ particles and denoted as $|N,l \ra$. Examples are the $W$ state $|W \ra = |3,1\ra = 1/\sqrt{3} (|001\ra + |010\ra + |100\ra)$ or $|4,2\ra = 1/\sqrt{6} (|0011\ra + |0110\ra + |1100\ra + |1001\ra + |1010\ra + |0101\ra)$.  Dicke states are detected by the witness \eqref{genentwit} with $c_x = c_y = 1, c_z =-1$. To see this, we calculate the expectation value $\la \Sigma(0)\ra$ of Eq.~\eqref{sigmagen} for the Dicke states $|N,l\ra$.
Entanglement is detected if $\la N,l | \Sigma(0)| N,l\ra >1$.
For the term $S^{zz}(0)$ we get $\la N,l | \hat{S}^{zz}(0) |N,l \ra = (4 \la J_z^2 \ra-N)/2 =
((N-2l)^2 -N)/2$.
Here we have used the fact that
the collective
spin operator $J_\alpha := 1/2 \sum_{k=1}^N \sigma^{\alpha}_k$ is given by
$\hat{S}^{\alpha \alpha}(0) = (4 J_\alpha^2 - N \id_N)/2$.
For $S^{xx}(0)$ and $S^{yy}(0)$ we obtain $\la N,l | \hat{S}^{xx}(0) |N,l \ra = \la N,l | \hat{S}^{yy}(0) |N,l \ra = l(N-l)$.
Since for the Dicke states we have $S^{xx}(0) = S^{yy}(0) \geq 0$ it is clear that the chances to detect entanglement are best for $c_x=c_y=1$.
For the case $c_x =c_y=1,c_z=-1$ we use the notation $\tilde{\Sigma}(0)$ and find
\begin{equation} \label{sigmank}
	\la N,l | \tilde{\Sigma}(0)| N,l\ra = \frac{4l(N-l)-(N-2l)^2+N}{N(N-1)}\,.
\end{equation}
In particular, for an even particle number $N$ and $l=N/2$ the expectation value \eqref{sigmank} becomes $\la \tilde{\Sigma}(0)\ra = (N+1)/(N-1) > 1$ and for odd $N$ and $l = (N-1)/2$ or $l=(N+1)/2$ we get $\la \tilde{\Sigma}(0)\ra = (N(N+1)-2)/N(N-1) > 1$, and thus these states are always detected. Also other Dicke states are detected, e.g. $|6,2\ra$ where $\la 6,2 | \tilde{\Sigma}(0)| 6,2 \ra = 17/15 \geq 1$.
Choosing different coefficients $c_{x,y,z}$ in the construction of the witness \eqref{genentwit}, other states can be detected. An interesting example for four particles is the superposition between two Greenberger-Horne-Zeilinger (GHZ) states, $(\cos{\theta})/2 (|0011\ra + |1100\ra) \pm (\sin{\theta})/2 (|0000\ra + |1111\ra)$. This state is detected for $\pi/4 < \theta < \pi/2$ if we choose $c_x =-1, c_y=c_z=1$ (minus sign) or $c_x=1, c_y=-1, c_z=1$ (plus sign). In Ref.~\cite{walther05} an experimental preparation of a four-qubit cluster state is reported, and from the presented method it seems likely that also the above GHZ superposition states can be prepared with this setup. Other detected symmetric states are superpositions of Dicke and GHZ states, e.g.,
for four particles $\cos{\theta} |4,2\ra \pm \sin{\theta}/\sqrt{2} (|0000\ra +|1111\ra)$ is detected for $\arccos{3\sqrt{2/19}} < \theta < \pi/2$ with the witness coefficients $c_x=-1,c_y=c_z=1$ (minus sign) and $c_x=1,c_y=-1,c_z=1$ (plus sign).

So far we have considered the case $k=0$ only. If we choose $k=\pi$ in the construction of the witness $W(k)$ \eqref{genentwit}, still more entangled states can be detected. Note that in this case the witness is no longer symmetric under particle exchange. An example of detected states are non-symmetric Dicke states with additional phases. Choosing $W(\pi)$ with $c_x=c_y=c_z=1$ we can detect the four-particle entangled state
\begin{align} \label{phaseddicke4}
	|D_4^{ph}\ra = \frac{1}{\sqrt{6}} \big( &|0011\ra + |1100\ra + |0110\ra + |1001\ra \nonumber\displaybreak[0]\\
	- &|0101\ra - |1010\ra \big).
\end{align}
For six particles and $W(\pi)$ with $c_x=c_y=c_z=1$ again, the state
\begin{alignat}{3} \label{phaseddicke6}
	|D_6^{ph}\ra &=& &\nonumber\displaybreak[0]\\
	\frac{1}{\sqrt{20}} &\big(& &|111000\ra + |001110\ra + |010101\ra + |011010\ra &\nonumber\displaybreak[0]\\
	&+& &|100011\ra + |100110\ra + |101001\ra + |101100\ra \nonumber\\
	&+& &|110010\ra + |001011\ra - |000111\ra - |110001\ra \nonumber\\
	&-& &|101010\ra - |100101\ra - |011100\ra - |011001\ra \nonumber\\
	&-& &|010110\ra - |010011\ra - |001101\ra - |110100\ra \big)
\end{alignat}
is detected. In general, all ``phased'' Dicke states $|N,l^{ph}\ra$, i.e. Dicke states with different signs before terms that correspond to even and odd permutations of $0$s and $1$s, and with $l=N/2$ for even N and $l=(N+1)/2$ or $l=(N-1)/2$ for odd N, are detected. Consider, e.g., the state $|6,3^{ph}\ra = |D^{ph}_6\ra$ \eqref{phaseddicke6}. Starting with the state $|111000\ra$, all even permutations, i.e. an even number of transpositions of neighboring spins, have a positive sign, and all odd permutations have a negative sign. This scheme can be generalized to construct the general phased Dicke states $|N,l^{ph}\ra$ of $N$ particles and $l$ excitations.

It is not necessary that all coefficients $c_{x,y,z}$ are non-vanishing. The states $|N,l^{ph}\ra$ for the mentioned particular values of $l$ are already detected by the witness $W(\pi)$ with $c_x=c_y=1, c_z=0$. This can be seen as follows: To determine $\la N,l^{ph} | \hat{S}^{xx}(\pi) |N,l^{ph}\ra$ and $\la N,l^{ph} | \hat{S}^{yy}(\pi) |N,l^{ph}\ra$ we note that we get exactly the same expressions as for the Dicke states, since the possible minus signs of the two-point correlation terms $\sigma^x_i \sigma^y_j$ and the minus signs of the phased Dicke states always exactly cancel out, due to the correspondence of an odd number of transpositions in the state and odd distance-terms $r_j - r_i$ in the structure factors. Thus for the phased Dicke states $|N,l^{ph}\ra$ we obtain $\la \hat{S}^{xx}(\pi) \ra $ = $\la \hat{S}^{yy}(\pi) \ra$ = $l(N-l)$, just as for the Dicke states $|N,l\ra$ in the previous paragraph. For phased Dicke states with even $N$ and $l=N/2$ and for $\Sigma(\pi)$ with $c_x=c_y=1, c_z=0$ we get $\la \Sigma(\pi) \ra = 2 \la \hat{S}^{xx}(\pi) \ra = N/(N-1) > 1$, and for odd $N$ with $l=(N+1)/2$ or $l=(N-1)/2$ we obtain $\la \Sigma(\pi) \ra = (N+1)/N > 1$. Therefore these states are always detected.

In order to prove that a given state carries genuine multipartite entanglement, one has to determine its minimal expectation value for any biseparable cut of the multi-qubit states. This task can in general only be performed numerically. Using the routines provided in Ref.~\cite{toth08}, we find the following results for the witnesses for phased Dicke states, when choosing $k=\pi$ and $c_x=c_y=c_z=1$: for 4 qubits, $\la \Sigma(\pi) \ra_{bisep} \leq 1.187$ and
for 6 qubits, $\la \Sigma(\pi) \ra_{bisep} \leq 1.158$, where these are upper bounds for all biseparable cuts. Comparing these numbers with the previous paragraph shows that there is a considerable range of genuine multipartite entangled states to be detected with our method.

The importance of the parameter $k$ for the detection of different types of states is remarkable:
The general phased Dicke states $|N,l^{ph}\ra$ that include the states \eqref{phaseddicke4} and \eqref{phaseddicke6}, are not detected by $W(0)$ for any choice of the coefficients $c_x$, $c_y$, and $c_z$, and $W(\pi)$ does not detect the ``usual'' Dicke states.

Furthermore, we want to study the robustness of the witness in Eq. \eqref{genentwit}
under the influence of noise. In particular, we consider two depolarizing channels; one that acts collectively on all qubits, which efficiently results in the addition of white noise to the multipartite state, and one that affects the single qubits \emph{independently}, i.e. that adds white noise to a single qubit $\rho_{s}; \ \rho_{s, dp} = (1-q) \rho_s + q \id/2$.

We study the collective depolarizing channel first. The Dicke states $|N,N/2 \ra$ then change accordingly to $p \id_N /2^N + \left(1-p\right) |N,N/2\ra \la N,N/2|$. The witness $W(0)$ with $c_x=c_y=1,c_z=-1$ detects entanglement of this state for $0 \leq p < 2/(N+1)$. The robustness decreases with a growing number of qubits $N$. In Ref.~\cite{toth05}, the robustness of certain witnesses against the collective depolarizing channel, corresponding to $c_z=0$, has been studied. These witnesses, that can detect Dicke states of the form $|N,N/2\rangle$, allow noise with $p<1/N$. Thus, adding the $z$-direction measurement to the witness improves its robustness against white noise. For the general ``phased'' Dicke states and using the witness $W(\pi)$ with $c_x=c_y=1$ and $c_z=0$, we obtain $0 \leq p < 1/N$ for entanglement detection. Again, a greater robustness can be achieved when an additional $z$-direction term, $c_z=1$, is introduced. For the noisy ``phased'' Dicke state (cf. Eq.~\eqref{phaseddicke4}), i.e. $p \id/16 +(1-p) |D_4^{ph}\ra \la D_4^{ph}|$, the witness $W(\pi)$ with $c_x=c_y=c_z=1$ detects entanglement of this state for $0 \leq p < 4/13$. In the case of six particles, it detects entanglement of the noisy state	$p \id /2^6 +(1-p) |D_6^{ph}\ra \la D_6^{ph}|$ (cf. Eq.~\eqref{phaseddicke6}) for a parameter interval $0 \leq p < 6/31$. The maximal values of $p$ are in both cases bigger than $1/N$.

In the following we investigate the robustness for the individual depolarizing channel,
where the noise affects each qubit independently. Since we are interested in expectation values only, it is
convenient to shift the influence of the Kraus operators $K_i$ characterizing the noise model to the observable
and leave the initial state unchanged.
This is possible, because in the operator sum representation of the
channel we have (where the subscript $dp$ denotes affection by the channel)
$\tr (O \rho_{dp}) = \sum_i \tr
( O K_i \rho K_i^\dag )= \sum_i \tr ( K_i^\dag O K_i \rho ) = \tr ( O_{dp} \rho )$.
Since the individual
depolarizing channel transforms the Pauli operators as $\sigma^\alpha_{dp} = (1-q) \sigma^\alpha, \ \alpha =
x,y,z$ (see, e.g., \cite{bruss00}), the two-point correlation terms of the structure factor simply change to $(\sigma^\alpha_i
\sigma^\beta_j)_{dp} = (1-q)^2 \sigma^\alpha_i \sigma^\beta_j$. Thus the observable $\Sigma(k)$ in
Eq.~\eqref{sigmagen} is influenced by the channel according to $\Sigma(k)_{dp} = (1-q)^2 \Sigma(k)$, and for the
expectation value we simply have $\la \Sigma(k)_{dp} \ra = (1-q)^2 \la \Sigma(k) \ra$. Thus we can determine the
robustness of $W(k)$, i.e. the region of $q$ for which $\la \Sigma(k)_{dp} \ra > 1$, where entanglement is
detected. For the Dicke states $|N, N/2 \ra$ and $c_x=c_y=1, c_z =-1$ for $W(0)$ we obtain a robustness region
$0 \leq q < 1-\sqrt{(N-1)/(N+1)}$. For the phased Dicke states $|N, N/2^{ph} \ra$ and $c_x=c_y=1, c_z=0$ for
$W(\pi)$ we find $0 \leq q < 1-\sqrt{(N-1)/N}$. For $c_x=c_y=c_z=1$ we calculated the four and six particle case
(see Eqs.~\eqref{phaseddicke4} and \eqref{phaseddicke6}), where we find $0 \leq q < 1- 3/\sqrt{13} \simeq 0.168$
for $|D^{ph}_4\ra$ and $0 \leq q < 1- 5/\sqrt{31} \simeq 0.102$ for $|D^{ph}_6\ra$. Both states exhibit a
robustness region that is larger for $c_z=1$ than for the $c_z = 0$ case, it therefore seems favorable to add the
z-direction measurement setting also in the case of the individual depolarizing channel.

While writing this paper, interesting experimental and theoretical articles on detecting
multi-qubit entanglement
of photonic Dicke states appeared online
\cite{prevedel09, wieczorek09,toth09, campbell09}.
The witnesses that were discussed and implemented there can be seen as specific cases
of the general construction presented here.
A detailed noise study
for witnesses of Dicke states that includes the amplitude and phase damping channels, in connection with the
numerical detection of genuine multipartite entanglement,
can be found in Ref.~\cite{campbell09}.
Note that a different method for macroscopic detection of non-separability,
which is valid for a certain class of spin models, is provided via the magnetic susceptibility, as shown in Ref.~\cite{wiesniak05}.

Summarizing, we present a general construction of entanglement witnesses for multi-qubit states and establish
a relation to static structure factors - these are macroscopic quantities which can be determined or measured in a
collective way in various periodic physical systems such as spin chains.
Our approach opens a wide avenue of possible
new connections for experimental detection of entanglement through scattering, and can
lead to a deeper understanding of entanglement properties in condensed matter systems.
Explicit constructions of witnesses for higher spins and eventually continuous variable
systems can follow along the same lines.
A connection to dynamical structure factors may allow also to study the dynamics of
entanglement.
Our method can be tested on spin chains with a finite number of sites, as for example in molecular magnetic
materials \cite{chen08}.

In suitable physical systems where the constituents can be addressed individually (such as polarized photons)
the witnesses can be measured locally.
Our witnesses are suitable for detecting a diversity of states, such as
Dicke states, non-symmetric versions of Dicke states, GHZ states and superpositions of these.
The presented construction also allows variations: in principle, arbitrary signs of the constituting two-point
correlations are possible.
This could be useful for the detection of various entangled states in photon experiments.
Thus, we offer more generality than previous two-point entanglement witnesses like spin squeezing inequalities
(see Ref.~\cite{guehne09} for an overview).
The question of whether the witnesses (that are constructed for a
finite number of subsystems) are also meaningful in the thermodynamical limit is still under investigation.

\emph{Acknowledgments}. We would like to acknowledge helpful discussions with Alessandro De Martino, and also with Sylvia Bratzik, Jens Eisert, Markus Mertz, Zahra Shadman, and Colin Wilmott. P. K. acknowledges financial support by FWF project CoQuS No. W1210-N16 of the Austrian Science Foundation. L. C. K. would like to acknowledge financial support by the National Research Foundation \& Ministry of Education, Singapore. This project was financially supported by the EU Integrated Project SCALA, the EU project CORNER, and by Deutsche Forschungsgemeinschaft (DFG). This work was supported in part by Perimeter Institute for Theoretical Physics.

\bibliography{refsmew1}

\end{document}